# A 10 Gbps Driver/Receiver ASIC and Optical Modules for Particle Physics Experiments

Xing Huang, Datao Gong, Suen Hou, Guangming Huang, Chonghan Liu, Tiankuan Liu, Ming Qi, Hanhan Sun, Quan Sun, Li Zhang, Wei Zhang, Xiandong Zhao, and Jingbo Ye

*Abstract*— We present the design and test results of a Drivers and Limiting AmplifierS ASIC operating at 10 Gbps (DLAS10) and three Miniature Optical Transmitter/Receiver/Transceiver modules (MTx+, MRx+, and MTRx+) based on DLAS10. DLAS10 can drive two Transmitter Optical Sub-Assemblies (TOSAs) of Vertical Cavity Surface Emitting Lasers (VCSELs), receive the signals from two Receiver Optical Sub-Assemblies (ROSAs) that have no embedded limiting amplifiers, or drive a VCSEL TOSA and receive the signal from a ROSA, respectively. Each channel of DLAS10 consists of an input Continuous Time Linear Equalizer (CTLE), a four-stage limiting amplifier (LA), and an output driver. The LA amplifies the signals of variable levels to a stable swing. The output driver drives VCSELs or impedance-controlled traces. DLAS10 is fabricated in a 65 nm CMOS technology. The die is 1 mm × 1 mm. DLAS10 is packaged in a 4 mm × 4 mm 24-pin quad-flat no-leads (QFN) package. DLAS10 has been tested in MTx+, MRx+, and MTRx+ modules. Both measured optical and electrical eye diagrams pass the 10 Gbps eye mask test. The input electrical sensitivity is 40 mVp-p, while the input optical sensitivity is -12 dBm. The total jitter of MRx+ is 29 ps (P-P) with a random jitter of 1.6 ps (RMS) and a deterministic jitter of 9.9 ps. Each MTx+/MTRx+ module consumes 82 mW/ch and 174 mW/ch, respectively.

## I. Introduction

HIGH-speed optical data transmission between on-detector and off-detector electronics is one of the key R&Ds in particle physics experiments [1]. Through common projects, CERN has demonstrated the lpGBT [2] serializer-deserializer (SerDes) and VTRx+ [3] optical transceiver up to 10.24 Gbps per fiber. The serializer in lpGBT operates at 5.12 and 10.24 Gbps, while its deserializer at 2.56 Gbps. VTRx+ has four transmitters and one receiver with fibers terminated at an MT connector. VTRx+ is only board-mount with a delicate electric connector. Both the lpGBT ASIC and the VTRx+ optical module have an ultra-small footprint and are qualified to be radiation tolerant for applications in the inner trackers of experiments on the LHC. Compared with the inner trackers, on-detector electronics for calorimeters and muon systems have much less stringent requirements in component size and radiation tolerance [4, 5]. The PCBs are also usually larger to house more functions such as the hardware trigger and high precision ADCs. The data channels and the PCB layout are more spread out, making it more natural to have a 1:1 transceiver and lpGBT match right after the ADC, and collect the digital data through fibers over such a large PCB to the front panel. The size and complexity of the PCB call for an optical transceiver with more robust connectors. Based on the development of the Miniature optical Transmitter/Transceiver MTx/ MTRx [6] and Vertical Cavity Surface Emitting Laser (VCSEL) driver LOCld [7] for the ATLAS Liquid Argon Calorimeter (LAr) trigger upgrade [8], we develop the Drivers and Limiting AmplifierS ASIC, operating at 10 Gbps (DLAS10), in a 65 nm CMOS technology and the corresponding three Miniature optical Transmitter/Receiver/Transceiver (MTx+, MRx+, and MTRx+). Comparing with its previous prototype of a dual-channel VCSEL driver [9], DLAS10 has two channels each works up to 10.24 Gbps and can be configured to be two VCSEL drivers, or two receiver limiting amplifiers, or one driver and one receiver. Matching DLAS10 with a TOSA and a ROSA with only Trans-Impedance Amplifier (TIA), and with a custom optical coupler, MTx+/MRx+/MTRx+ offer an economical option with a robust electrical connector and receives fibers with the LC connectors [10]. The three variants, MTRx+, MTx+, and MRx+, all stay below 6 mm in height and are both board and panel mountable, cover the needs of optical transceiver, dual optical transmitter and receiver, for on-detector readout electronics outside the inner trackers.

## II. Design of DLAS10

### A. Structure of DLAS10

Each channel of DLAS10 consists of an input buffer, a four-stage Limiting Amplifier (LA), and an output driver. Two channels share an Inter-Integrated Circuit (I$^2$C) block. Fig. 1 shows that the block diagram of DLAS10 drives two TOSAs,

This work was supported by SMU's Dedman Dean's Research Council Grant.

Xing Huang, Hanhan Sun, Li Zhang, and Wei Zhang are PhD candidates of Central China Normal University, Wuhan, Hubei 430079, China. They are visiting scholars at Southern Methodist University, Dallas, TX 75205, USA for this work.

Datao Gong (corresponding author, e-mail: dgong@mail.smu.edu), Chonghan Liu, Tiankuan Liu, and Jingbo Ye are with Southern Methodist University, Dallas, TX 75205, USA.

Suen Hou is with Academia Sinica, Nangang, Taipei 11529, Taiwan.

Guangming Huang is with Central China Normal University, Wuhan, Hubei 430079, P.R. China.

Ming Qi is with Nanjing University, Nanjing, Jiangsu 210008, China.

Quan Sun is with Fermi National Accelerator Laboratory, Batavia, IL 60510, USA.

Xiandong Zhao was with Southern Methodist University, Dallas, TX 75205, USA. He is now with Washington University in St. Louis, 1 Brookings Dr, St. Louis, MO 63130, USA.

receives the signal from a ROSAs with only TIA, connects to a TOSA a ROSA. DLAS10 is powered by a single 1.2 V power supply, whereas TOSAs and ROSAs are powered by a 3.3 V power supply. DLAS10 is AC coupled to TOSAs and ROSAs.

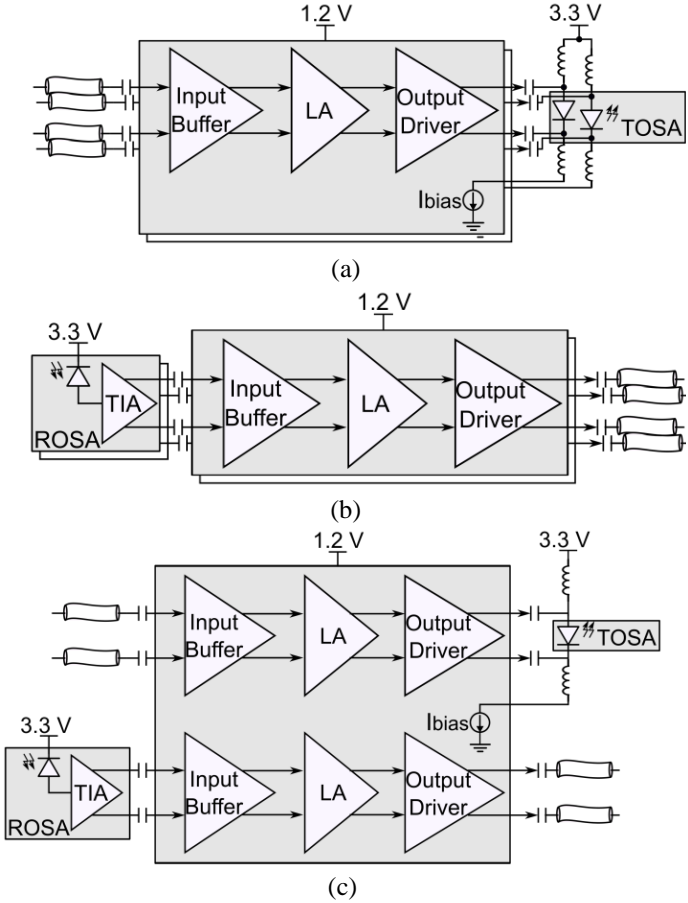

Fig. 1. Block diagrams of DLAS10 in the use case of MTx+ (a), MRx+ (b), and MTRx+ (c).

### B. Input buffer

The schematic of the input buffer is shown in Fig. 2. To cope with the cases where the input signal may suffer from high-frequency loss through long transmission lines, an inductive peaking technique [10] is used to boost signals at high frequencies. The conventional Continuous-Time Linear Equalization (CTLE) [12] with source degeneration using a combination of a capacitor and a variable resistor is also implemented to adjust the locations of the zeroes and poles in the frequency response to further compensate for the high-frequency degradation of the input signal. The tunable resistor is implemented by 6 pairs of resistors and NMOS transistors in series. The minimum input signal level is designed to match the output of a typical TIA from a ROSA. The CTLE is programmable and can compensate signal loss at 5 GHz up to 9.8 dB.

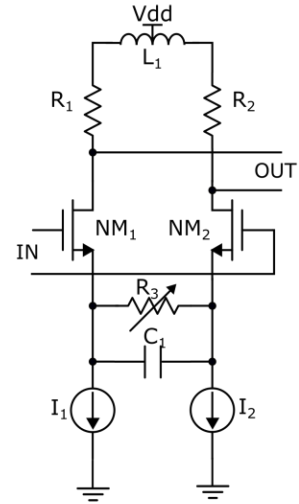

Fig. 2. Schematics of the input buffer with inductive peaking and CTLE.

### C. Limiting Amplifier

The LA aims at amplifying a wide range of input signals to the saturation level and to the output driver. The schematic of the LA is shown in Fig. 3. To obtain the overall sensitivity, a gain of 24 dB is required. In addition to achieving the gain, maintaining a wide bandwidth at all stages is also a key feature to avoid the Inter-Symbol Interference (ISI). A multiple-stage LA with an inductance-peaking technique is adopted [13]. Center-tapped inductors, shared between adjacent amplifier stages, are used to save layout area. An inductor is shared between adjacent amplifier stages for further area saving [14] To accommodate the Processes, Voltages and Temperatures (PVT) variations, an active feedback is used [15]. The active feedback transconductance can be tuned by the I$^2$C block to obtain a reasonable combination of bandwidth and gain. The simulated bandwidth and gain for LA are 10.5 GHz and 28.4 dB.

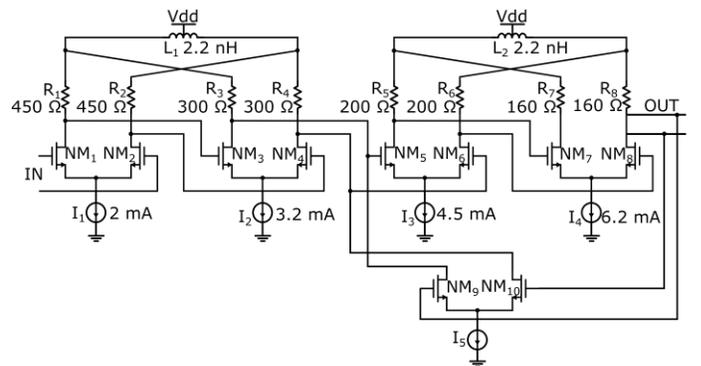

Fig. 3. Schematic of the four-stage LA.

### D. Output driver

The schematic of the output driver is shown in Fig. 4. NM$_1$ and NM$_2$ are the amplifying transistors performing as current switches. NM$_3$ and NM$_4$ form a current mirror. The current source I$_1$ is a simplified representation of the digital-to-analog converter (DAC) and is configured via the I$^2$C block to adjust the output modulation amplitude for applications as a cable driver or a VCSEL driver. The output driver is based on Current-Mode Logic (CML) structure and has 50 Ω load resistors to match the transmission-line impedance on the Printed Circuit Board (PCB). A large current capability is required to achieve a sufficient output swing.

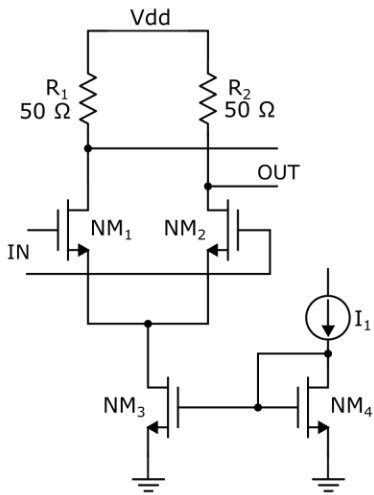

Fig. 4. Schematics of the output driver.

*E. Dimension and package*

The die of DLAS10 is 1 mm × 1 mm and is packaged in a 4 mm × 4 mm 24-pin Quad-Flat No-leads (QFN) chip. Microphotograph of the die and picture of the packaged chip are shown in Fig. 5(a) and Fig. 5(b), respectively.

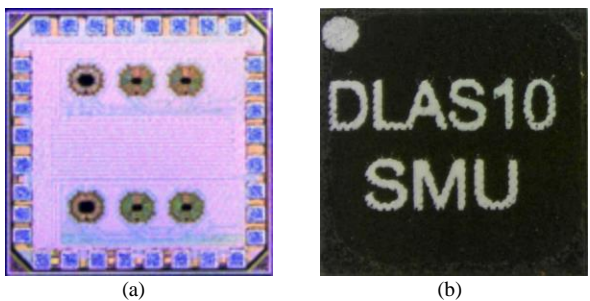

Fig. 5. (a) Microphotograph of the die; (b) picture of the packaged chip.

## III. Design of Optical Modules

DLAS10 has been tested in the optical transmitter/receiver (MTx+, MRx+, and MTRx+) modules. These modules are assembled with the custom mechanical latch that couples the module and optical fibers with LC connectors. Fig. 6(a) shows the MTx+ and MRx+ modules compared with a quarter dollar coin. Fig. 6(b) shows the test board with MRx+ mounted. The electric connector is that for the SFP+ modules. The latch can be anchored to the motherboard when the module is used as board mount. It also has an EMI shield that doubles as a plug-in guide when the module is used at a front-panel. The part numbers of the TOSAs and the ROSAs used in the test are Truelight TTF-1F59-427 and TRF-8F59-732, respectively.

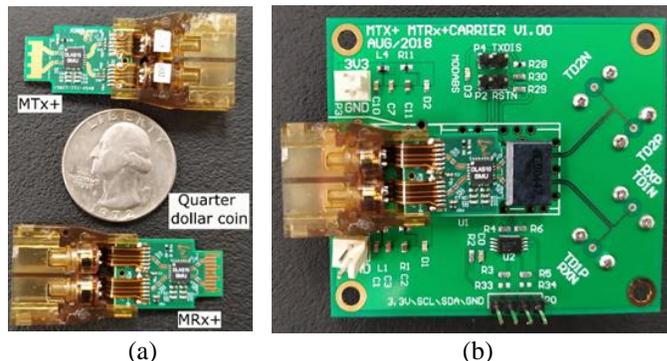

Fig. 6. (a) Photos of MTx+ and MRx+ compared with a quarter dollar coin; (b) the carrier board with an MRx+ mounted.

## IV. Test Results

*A. Test setup*

Fig. 7 (a) and Fig. 7(b) show the test block diagrams of MTx+ and MRx+, respectively. Fig. 6(c) shows the picture of the test setup. Two DC power supplies (Agilent E3641A) provide 1.2 V and 3.3 V power. A pattern generator (CENTELLAX TG1C1-A with a clock module CENTELLAX PCB12500) provides 10 Gbps Pseudo-Random Binary Sequence (PRBS) signals with various amplitudes and the trigger clock. An optical oscilloscope (Tektronix TDS8000B) captures the optical eye diagrams of MTx+. An electrical oscilloscope (Tektronix DSA72004) measures the electrical eye diagrams of MRx+. For the MRx+ test, MTx+ is the optical source.

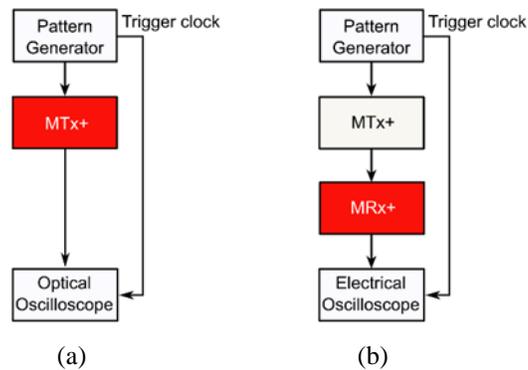

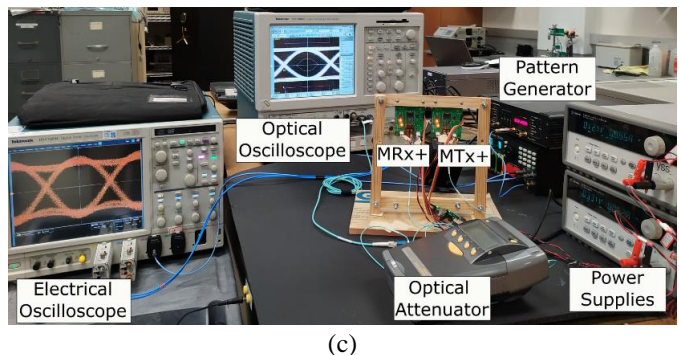

Fig. 7. (a) and (b) Test block diagrams of MTx+ and MRx+, (c) photo of the test.

*B. Input sensitivity*

The input sensitivities of MTx+ and MRx+ were tested by scanning the input signal amplitude. Fig. 8 shows the input sensitivities of MTx+ and MRx+, respectively. As can be seen in Fig. 8(a), when the input amplitude exceeds 40 mV, the Optical Modulation Amplitude (OMA) of MTx+ is stable. The input sensitivity of 40 mV is low enough to receive the signal from the ROSA that has no LA. As can be seen in Fig. 8(b), when the input OMA is higher than -12 dBm, the bit error rate (BER) is below $1\times10^{-12}$. The optical input sensitivity of -12 dBm is comparable with commercial optical transceiver modules.

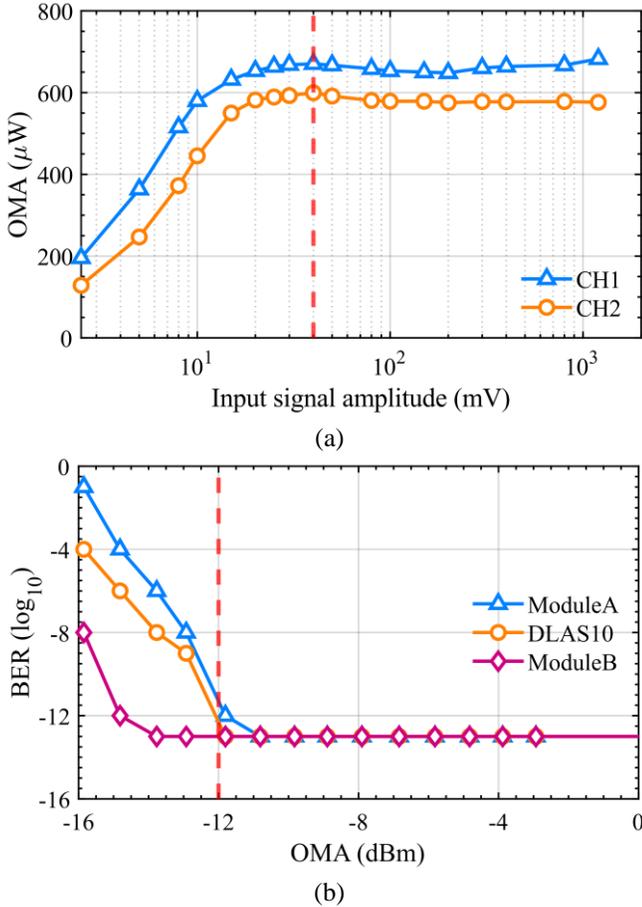

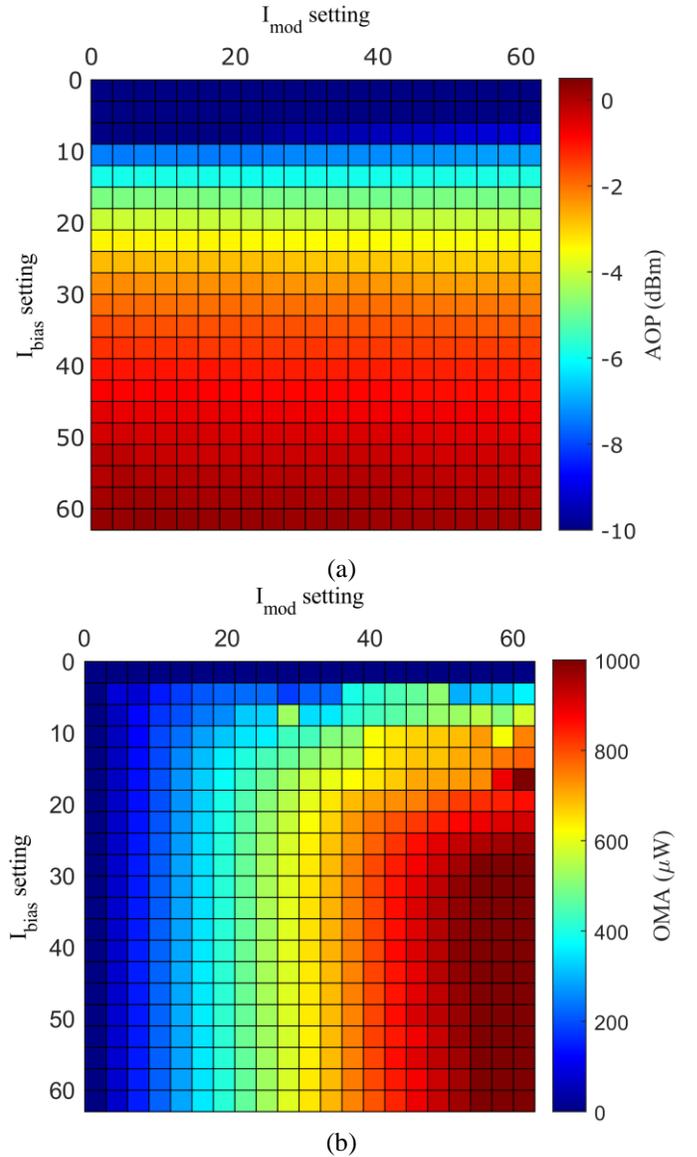

Fig. 8. (a) OMA versus the amplitude of the input electrical signal of an MTx+; (b) BER versus the OMA of MRx+, compared with two commercial modules.

Fig. 9. (a) AOP versus $I_{mod}$ and $I_{bias}$ joint scanning; (b) OMA versus Imod and Ibias joint scanning.

*C. $I_{mod}$ and $I_{bias}$ scanning of MTx+*

The modulation current ($I_{mod}$) and the bias current ($I_{bias}$) of MTx+ were scanned for the optimal operational parameters. The Average Output Power (AOP) and OMA of the test were recorded in the 2-dimension plots shown in Fig. 9(a) and Fig. 9(b), respectively. The AOP is strongly correlated to the $I_{bias}$ setting but not the $I_{mod}$ setting. When $I_{bias}$ is small, the correlation between OMA and Imod is not obvious. When the $I_{bias}$ setting reaches a certain level, OMA is positively correlated with the $I_{mod}$ setting. The $I_{bias}$ setting of 48 and the $I_{mod}$ setting of 32 were selected for the following test of MTx+, according to the results of the $I_{bias}$ and $I_{mod}$ scanning.

*D. Input equalization of MTx+*

The effect of the input equalization was tested by scanning the input equalization setting using two 2 m long coaxial cables with Sub-Miniature version A (SMA) connectors. The results are shown in Fig. 10(a). The RMS jitter can be optimized by adjusting the input equalization setting. When the input equalization setting is 62, the RMS jitter achieves 2.8 ps, which alleviates the high frequency attenuation caused by the long input transmission line. The eye diagram is shown in Fig. 10(b) when input equalization setting is 0. The eye diagram is shown in Fig. 10(c) when input equalization setting is 62.

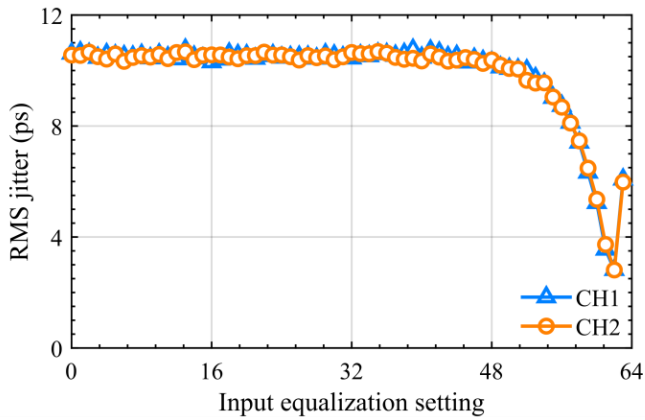

(a)

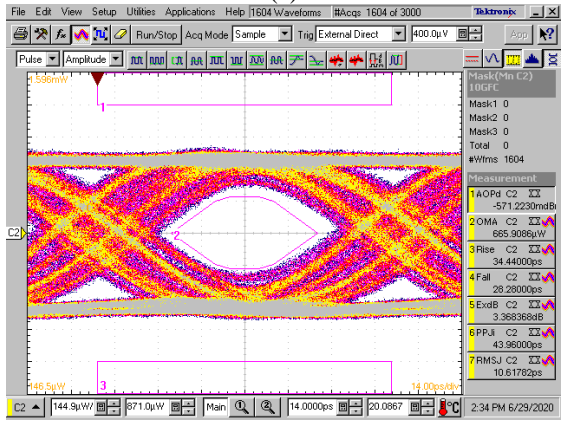

(b)

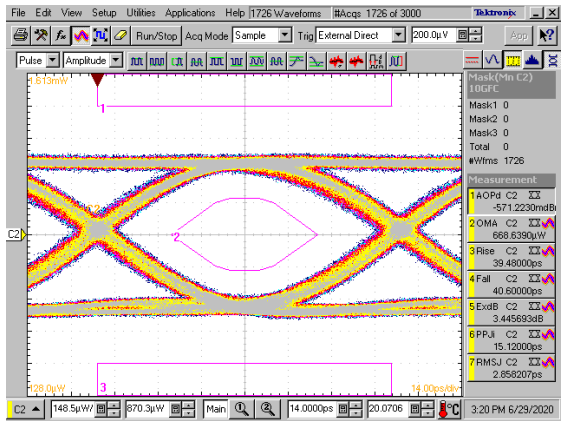

(c)

Fig. 10. (a) Effects of the input equalization; (b) eye diagram when the input equalization setting is 0; (c) eye diagram when the input equalization setting is 62.

### E. Feedback of LA

The goal of the LA feedback is to adjust the balance between the gain and bandwidth of the LA. When the feedback is strong, the LA has high bandwidth and low gain, and vice versa. A relatively small input can emphasize the relationship between feedback and gain. 10 mV is selected as the input amplitude for testing the feedback of LA.

Fig. 11 shows the effects of the feedback strength (FBS). OMA decreases when FBS setting increases because of the gain of LA decreases. In the meantime, the bandwidth of LA increases, reflecting the decrease in rise and fall times.

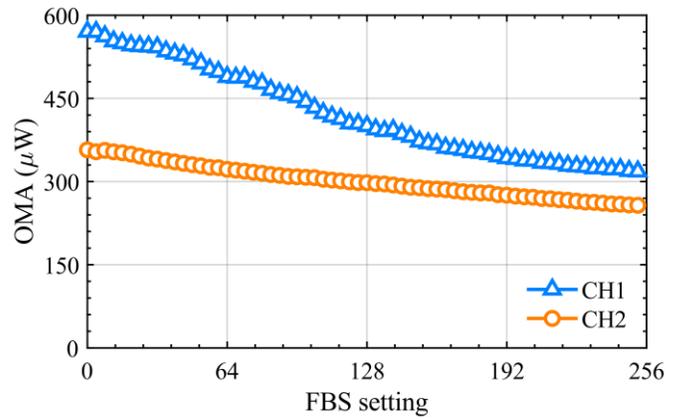

(a)

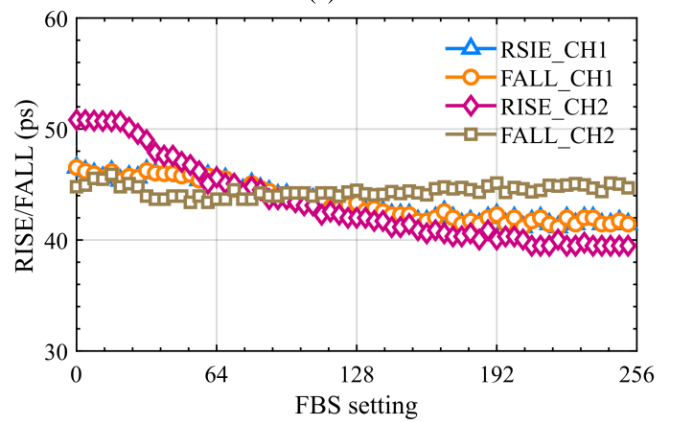

(b)

Fig. 11. (a) Dependence of the OMA and (b) the transition times on the FBS.

### F. Eye diagrams

The eye diagrams of MTx+ and MRx+ are shown in Fig. 12(a) and Fig. 12(b), respectively. Both eye diagrams passed the 10 Gbps eye mask test.

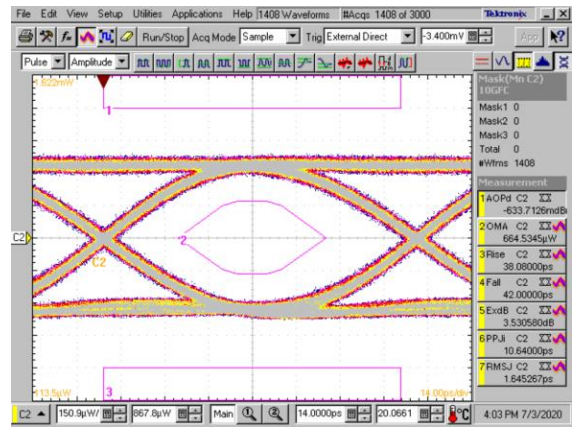

(a)

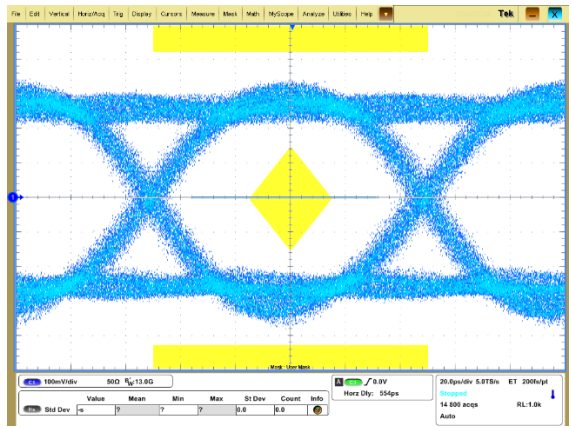

(b)

Fig. 12. (a) Eye diagrams of MTx+ and (b) MRx+.

### G. Jitter measurements

The jitter measurement results of MTx+ are shown in Fig. 12 (a). Although no decomposition of random jitter and deterministic jitter is performed by the optical scope, the measured peak-to-peak jitter is 10.6 ps and RMS jitter 1.6 ps. The jitter of MRx+ is analyzed with the electrical oscilloscope. As can be seen in Fig. 13, the total jitter of MRx+ is 29 ps (P-P) with a random jitter of 1.6 ps (RMS) and a deterministic jitter of 9.9 ps (P-P). This should be understood as the parameter taken with MTx+ as the source input.

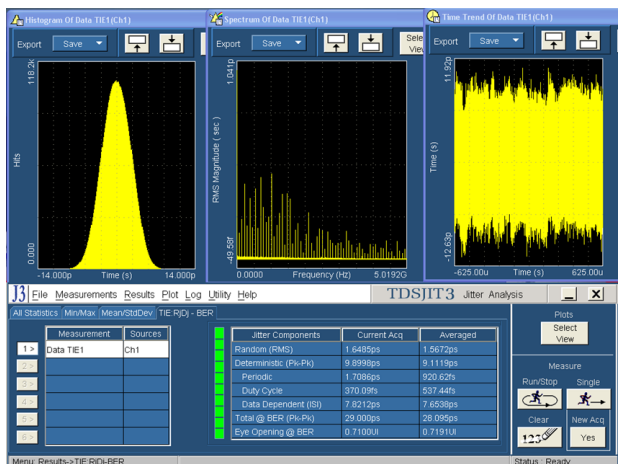

Fig. 13. Jitter measurement of MRx+.

### H. Power consumption

Each MTx+/MTRx+ module consumes 82 mW/ch and 174 mW/ch, respectively.

### I. Radiation tolerance

The radiation tolerance of DLAS10 and the ROSAs used in the MRx+/MTRx+ modules will be tested in the future. The previous prototype of DLAS10 and the TOSAs that are used in MTx+/MTRx+ have been verified to be radiation tolerant for applications in certain particle physics experiments [9, 16].

## V. CONCLUSION

We have designed and tested a 10 Gbps driver/receiver ASIC DLAS10 and Optical Transmitter/Receiver/Transceiver modules MTx+, MRx+, and MTRx+ that are based on DLAS10. DLAS10 can drive TOSAs and receive signals from ROSAs with only a TIA. DLAS10 is fabricated in a 65 nm CMOS process. The die of DLAS10 is 1 mm × 1 mm. The electrical input sensitivity of MTx+ is 40 mVpp. The optical input sensitivity of MRx+ is -12 dBm. The power consumptions of MTx+ and MRx+ are 82 mW/ch and 174 mW/ch, respectively.

### ACKNOWLEDGMENT

We thank Szymon Kulis and Paulo Moreira at CERN for sharing the I²C block design. We also thank financial support from the SMU University Research Council Grant, the Dedman College Dean's Research Council Grant, and from the Institute of Physics, Academia Sinica.